\documentclass[a4paper,11pt]{article}
\usepackage{jheppub}
\usepackage{lineno}
\usepackage{amsmath}

\DeclareMathOperator{\arccot}{arccot}

\newcommand{\be}{\begin{equation}}
\newcommand{\ee}{\end{equation}}

\title{AdS$_3$ axion wormholes as stable contributions to the Euclidean gravitational path integral }

\author[a]{Andrew Loveridge,}
\affiliation[a]{Department of Physics, the University of Texas at Austin,\\
2515 Speedway, Austin, TX 78712, USA}
\emailAdd{aloveridge@utexas.edu}

\author[b]{Hao-Yu Sun}
\affiliation[b]{Weinberg Institute, the University of Texas at Austin,\\
2515 Speedway, Austin, TX 78712, USA}
\emailAdd{hkdavidsun@utexas.edu}

\abstract{
Recent work has demonstrated that Euclidean Giddings--Strominger axion wormholes are stable in asymptotically flat 4D Minkowski spacetime, suggesting that they should, at least naively, be included as contributions in the quantum gravitational path integral. Such inclusion appears to lead to known wormhole paradoxes, such as the factorization problem. In this paper, we generalize these results to AdS$_3$ spacetime, where the axion is equivalent to a $U(1)$ gauge field. We explicitly construct the classical wormhole solutions, show their regularity and stability, and compute their actions for arbitrary ratios of the wormhole mouth radius $\tau_{\text{min}}$ to the AdS radius $l$ and across various topologies. Finally, We discuss potential implications of these findings for the 3D gravitational path integral.
}

\begin{document}
\maketitle
\flushbottom

\section{Introduction}
\label{sec:intro}

On theoretical grounds, one expects that a quantum theory of gravity should incorporate contributions from all possible spacetime topologies consistent with given constraints, boundary conditions, and initial or final data. This enhanced background independence \cite{Smolin:2005mq} is compelling because it could elucidate various topological properties of our universe, such as  spacetime dimensionality, connectedness, and manifold structure.

Indeed, incorporating diverse contributions in Euclidean quantum gravitational path integrals successfully reproduces black hole thermodynamics \cite{HawkingPage, Witten:1998zw} and the Page curve associated with black hole evaporation \cite{Penington:2019kki, Almheiri:2019hni}, suggesting intriguing connections to unitarity and holography \cite{Rey:1998yx}. Similarly, in Lorentzian signature, string theory employs a summation over topologies as a worldsheet quantum gravity theory, and allows for topological changes even within the target space \cite{Aspinwall:1993nu, Witten:1996qb, Brandle:2002fa}.

On the other hand, it has also long been known that if topologically nontrivial spacetimes with wormholes are included in the gravitational path integral, new pathologies can emerge \cite{Stephens:1993an, tHooft:1993dmi, Coleman:1988cy, Giddings:1988cx, Giddings:1988wv} (for a nice review, see \cite{Hebecker:2018ofv}). For example, single boundary wormholes can violate the \textit{cluster decomposition principle} \cite{Arkani-Hamed:2007cpn}, thus undermining locality, or induce random couplings \cite{Coleman:1988cy, Giddings:1988cx}; multiboundary wormholes lead to the factorization problem \cite{Maldacena:2004rf, Witten:1999xp} or necessitate disorder averaging \cite{Marolf:2020xie}. Consequently, a fundamental tension emerges between locality, background independence, and unitarity in quantum gravity.

Several possibilities exist to resolve this tension. Perhaps quantum gravity does not admit topology changes, rendering previous results misleading.  Alternatively, many or all apparently topological distinct spacetimes might be related to each other \cite{Eberhardt:2021jvj}, or additional geometric or non-geometric contributions could be required. There are a variety of other viable and promising proposals \cite{Saad:2021rcu, Saad:2021uzi, Iliesiu:2021are, Schlenker:2022dyo, Hernandez-Cuenca:2024pey}. However, the most conservative solution we may hope for, as explained by \cite{Hertog:2018kbz}, would be that the pathology-inducing topologies are simply unstable, eliminating their contributions from the saddle-point approximation of the gravitational path integral, while any ``useful'' topologies are instability-free.

Hertog et al. \cite{Hertog:2018kbz} examined the stability of the classic Euclidean Giddings--Strominger axion-sourced wormhole \cite{Giddings:1987cg}, which prompted much of the classic work on wormhole-related apparent paradoxes. To a sigh of relief, they appeared to show that the solutions were simply unstable, at least in $4$D asymptotically flat spacetime or AdS$_4$ with a large AdS radius relative to the wormhole mouth radius. However, subsequent analyses \cite{Loges:2022nuw} revealed that in the dual description of the axion as a $2$-form field strength\footnote{While this manuscript was undergoing revision, the very relevant \cite{witten2026dualityaxionwormholes} appeared, offering a systematic assessment of this duality of descriptions and associated subtleties.},
 the wormholes are actually stable in flat space. The source of the discrepancy, as fully elucidated later in \cite{Hertog:2024nys}, is a difference in boundary conditions: authors of 
\cite{Hertog:2018kbz} implicitly chose a Robin boundary condition. For boundary conditions which fix the axion charge and boundary metric, it is now clear from \cite{Loges:2022nuw, Hertog:2024nys} that the axion wormholes are stable, in spite of other boundary conditions leading to instability.

Still, one may expect that in string theory, additional effects from other fields could render the solutions unstable. An especially important case, theoretically speaking, is for an AdS background, since here one is seemingly led to violations of cluster decomposition in the dual CFT. This is not covered by the results of \cite{Loges:2022nuw}, and was examined only perturbatively in \cite{Hertog:2018kbz} due to computational intractability in $4$D.

In recent years much progress has been made by considering toy models of quantum gravity in dimensions $D \le 3$. Here we are interested in $D=3$, see e.g., \cite{Maloney:2007ud,Castro:2011zq,Jian:2019ubz,Karch:2020flx, Cotler:2020ugk, Cotler:2020hgz, Collier:2023fwi, Chandra:2022bqq, Collier:2024mgv}. While these models lack propagating gravitons—a significant limitation—they nonetheless feature black holes \cite{Banados:1992gq, Mertens:2022irh} and capture other essential gravitational phenomena, offering valuable computational tractability.

In $3$D, the axion field responsible for Giddings--Strominger wormholes is equivalent to a $U(1)$ gauge field, the “dual photon” \cite{TongGaugeTheory}.
\footnote{Although, importantly, the boundary conditions must be chosen differently in $D=3$, see section \ref{BCandUV}}.
This provides an especially tractable situation with well-studied holographic applications \cite{Marolf_2006}, enabling detailed analysis of wormhole contributions and associated paradoxes. That is precisely the aim of this paper, where we extend previous results \cite{Hertog:2018kbz,Loges:2022nuw, Hertog:2024nys,Barcelo:1995gz} to AdS$_3$ backgrounds, for arbitrary ratio of the wormhole mouth radius to the AdS radius, and to torus and compact hyperbolic topologies, placing wormholes in a context whose implications may be more theoretically tractable.

This paper is organized as follows: In Section \ref{sec:solution}, we derive the classical solutions and demonstrate their regularity. In Section \ref{stability}, we show that the solutions are stable. In Section \ref{action}, we compute the action and compare it with other solutions with identical boundary conditions. Finally in Section \ref{discussion}, we discuss potential implications for 3D quantum gravity, the AdS$_3$/CFT$_2$ correspondences, and string theory, and recommend future work.

\section{Classical solutions}
\label{sec:solution}
In this section we find classical solutions which are candidate saddle point contributions to the Euclidean quantum gravity path integral with wormhole topologies.

\subsection{Euclidean action}
We consider Euclidean Einstein--Hilbert quantum gravity in $D=3$ dimensions with a $U(1)$ gauge field and a negative cosmological constant $\Lambda=-1/l^2$, where $l$ is the AdS radius. Due to the nature of our ansatz and apsects of the differential topology of $D=3$, the ADM formalism \cite{Arnowitt:1959ah} is convenient. The gravitational action in ADM form with choice of shift vector $N^i=0$ in Lorentzian signature is
\be
S_g=\frac{1}{2 \kappa} \int N \sqrt{h}(R_h-2\Lambda +K_{ab} K^{ab}-K^2),
\ee
where $h$ and $R_h$ refer to the spatial 2-metric and
\be
\label{Kab}
K_{ab}= \frac{1}{2 N} \dot{h}_{ab},
\ee
with the dot denoting a derivative with respect to time. We can Euclideanize the action by replacing $N\to - i N$ to get
\be
I_g=-\frac{1}{2 \kappa} \int N \sqrt{h}\big(R_h-2\Lambda -(K_{ab} K^{ab}-K^2) \big),
\ee
where $K_{ab}$ retains the form \eqref{Kab}. We may choose the lapse function to depend only on Euclideanized time $\tau$:
\be
N \equiv N(\tau).
\ee
If we do this, then as is well known the curvature term becomes a total derivative. In particular the Chern--Gauss--Bonnet theorem states that
\be
\int \sqrt{h} R_h= 4 \pi \chi,
\ee
where $\chi$ is the Euler characteristic of the spatial 2-manifold. This means the gravitational action can be reduced to
\be
\label{ADMaction2}
I_g= -\frac{4 \pi \chi}{2 \kappa} \int d\tau N+\frac{1}{2 \kappa} \int d\tau dV N \sqrt{h}(K_{ab} K^{ab}-K^2+2 \Lambda).
\ee
The Euclideanized $U(1)$ (electromagnetic) action is
\be
\label{emaction1}
I_{em}= \int \sqrt{g} \frac{1}{4 \alpha } F_{ab}.F^{ab},
\ee
where $g$ is the full $D=3$ metric. If we employ the fact that the shift vector is zero we can decompose this into
\be
\label{EMaction}
I_{em}= \int N \sqrt{h} \left( \frac{1}{2 \alpha N^2} E_i E^i+\frac{1}{2 \alpha g} B^2\right),
\ee
where $E_i=F_{0i}$ and $B=F_{12}$.

As is well known, while there are no propagating bulk gravitational degrees of freedom, there are gravitons associated with the boundary, and other nontrivial gravitational effects such as black holes are possible. The $U(1)$ field is special in $D=3$ since it is equivalent to a ``dual photon'' scalar field, which is an axion. We will stick with the $U(1)$ description, following \cite{Loges:2022nuw, Hertog:2024nys}.

For reference, it is worth noting that in $D=3$, $\kappa$ has dimensions of length , while $\alpha$ has units of inverse length. The latter would make the $U(1)$ field confining in the presence of charged matter \cite{TongGaugeTheory}, which does not matter for our considerations here but could affect generalizations or string embeddings which include other fields.

\subsection{Spherical topology}
We would like to find the $D=3$ analogue of the Giddings--Strominger wormhole \cite{Giddings:1987cg}, which would mean topology $R^1\times S^2$. Our ansatz is
\be
\label{sphereansatz}
ds^2= N_0^2 d\tau^2+\tau^2(d\phi^2+\sin^2\phi d\theta^2), \quad B= \frac{n}{4 \pi} \sin{\phi}, \quad E=0.
\ee
The choice of $B$ reflects charge quantization:
\be
\int_{S^2} F=n.
\ee
The minisuperspace action becomes (after integrating over the sphere)
\be
I_{total}=\int d\tau \left(-\frac{4 \pi  \left(l^2+N^2 \tau^2\right)}{\kappa  l^2 N}+\frac{n^2 N}{8 \pi \alpha \tau^2}-\frac{2 \pi N \chi }{\kappa } \right),
\ee
where $\chi=2$ for the sphere. The Hamiltonian constraint gives the solution:
\be
\mathcal{H}=\frac{\partial L}{\partial N_0}=\frac{4 \pi}{\kappa}\left(\frac{1}{N_0^2}-\frac{\tau^2}{l^2}-1\right)+\frac{n^2}{8 \pi \alpha \tau^2 }=0 \implies
N_0=\frac{1}{\sqrt{1+\frac{\tau^2}{l^2}-\frac{\kappa  n^2}{32 \pi ^2 \tau^2 \alpha}}}.
\ee
The momentum constraint states that
\be
\nabla^i ( K_{ij}-Kg_{ij})=0.
\ee
But the quantity in the parentheses above is equal to $-g_{ij}/(\tau N)$ and so is zero automatically since $\nabla^i$ involves only spatial derivatives.

Satisfying the Hamiltonian and momentum constraints is equivalent to satisfying the full Einstein equations, so this is a legitimate solution to the full equations of motion.

The Gauss's law constraint, since there is no charge, requires
\be
d \star F=0.
\ee
Indeed we have
\be
d \star F=d\left( \frac{n}{4 \pi} N d\tau\right)=0,
\ee
since $N$ depends only on $\tau$.

\begin{figure}[h]
    \centering
    \includegraphics[scale=1]{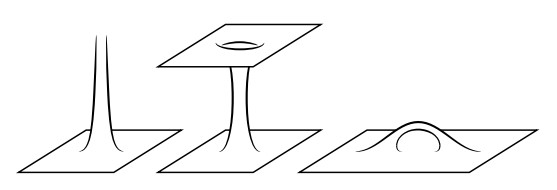}
    \caption{Three wormhole geometries considered in this paper. The first is the half wormhole solution found as a solution of the Euclidean equations of motion which has a minimal value of $\tau$, and can be thought of as related to the amplitude for emission of a ``baby universe''. The second is a two boundary wormhole made from ``gluing'' two half solutions together. The third is a single boundary wormhole which comes from identifying the boundaries of the two boundary solution, which is for large mouth separation asymptotically a solution to the equation of motion with the same action as the two boundary solution, in the sense of an instanton gas approximation. All of this is explained nicely in the review \cite{Hebecker:2018ofv} from which we borrowed this illustration. }
    \label{wormholeshapes}
\end{figure}

This is therefore a solution to the equations of motion. However, there is a (coordinate) singularity at the minimal value of $\tau$:
\be
\label{tauminsphere}
\frac{\tau_{\text{min}}^2}{l^2}= \sqrt{ \frac{\kappa n^2}{32 \pi^2 \alpha l^2 } +\frac{1}{4} }-\frac{1}{2} \ge 0.
\ee
This minimal value of $\tau$ reveals we have found only half of our wormhole solution. We must ``glue'' two copies of it together to get the complete solution. See figure \ref{wormholeshapes} for a visual provided by the nice review \cite{Hebecker:2018ofv}. In so doing we must ensure the spacetime remains a smooth solution to the euclidean equations of motion.

After some experimentation, one can see a convenient method of gluing turns out to be to switch to a new coordinate defined by
\be
\tau^2=T^2+\tau_{\text{min}}^2,
\ee
in which the metric solution becomes
\be
\label{newmetric}
ds^2= \frac{1}{1+T^2/l^2+2 \tau_{\min}^2/l^2} d T^2+(T^2+\tau_{\min}^2)(d \phi^2+\sin^2\phi d \theta ^2 ).
\ee
Notice this no longer suffers from any awkward coordinate singularity and $T$ can be extended to run from $-\infty$ to $\infty$ in an entirely symmetric way. The solution remains regular, with e.g., the Kretschmann scalar characterizing the curvature of $g_{ab}$ obeying:
\be
\frac{12}{l^4} \le R_{abcd}R^{abcd} \le \frac{32}{l^4}+\frac{32}{l^2 \tau_{\text{min}}^2}+\frac{12}{\tau_{\text{min}}^4},
\ee
with the minimum obtained at $T \to \pm \infty$ and the maximum at the wormhole ``throat'' $T=0$. This is finite as long as the throat itself has nonzero radial size $\tau_{\text{min}}>0$ which is true as long a there is \emph{some} magnetic (axion) charge $n>0$.

Meanwhile the $U(1)$ field strength obeys:
\be
\label{fieldstrengthsphere}
0\le F_{ab}F^{ab}= \frac{n^2}{8 \pi^2 (T^2+\tau_{\text{min}}^2)^2} \le \frac{n^2}{8 \pi^2 \tau_{\text{min}}^4},
\ee
which similarly is nonsingular.

One can check that this extended metric ansatz \eqref{newmetric} with the field strength from \eqref{sphereansatz} solves the full Einstein equation as long as $\tau_{\text{min}}$ obeys \eqref{tauminsphere}, so this is a complete and non-singular two-sided solution.

\subsection{Torus topology}

It is natural to try to generalize the preceding solution to wormholes of other topologies. Due to its connection with finite temperature effects in a hypothetical dual field theory, the torus topology is of particular interest. Using the same procedure as before, we find:
\be
\label{TorusSolutionMetric}
ds^2= \left(\frac{\tau^2}{l^2}-\frac{\kappa n^2 }{2 V^2 \alpha \tau ^2}\right)^{-1}d\tau^2+\tau^2(dx^2+dy^2),
\ee
\be
F= \frac{n}{V} dx \wedge dy,
\ee
where $V=\int dx \wedge dy$ is the torus coordinate volume, with $x$ and $y$ dimensionless and periodically identified, is a solution to the equations of motion. For $n\ne 0$ this is gives a smooth wormhole geometry. If $n\to 0$ the result is not AdS$_3$ but a geometry with a cusp singularity at $\tau=0$. In fact taking this limit is a nice illustration of the way a naive periodic identification of \eqref{TorusSolutionMetric} with $n=0$ does not actually give vacuum AdS$_3$, which was noted in \cite{Maloney:2007ud}. 

This solution also has a coordinate singularity at a minimum value of $\tau$:
\be
\frac{\tau_{\text{min}}^2}{l^2}= \sqrt{\frac{\kappa n^2}{2 V^2 \alpha l^2}}.
\ee
Once again we may extend the solution. One finds that
\be
\label{newmetrictorus}
ds^2= \frac{1}{T^2/l^2+2 \tau_{\min}^2/l^2} d T^2+(T^2+\tau_{\min}^2)(dx^2+dy ^2 )
\ee
provides a complete two-sided solution, and is smooth for $n>0$:
\be
\frac{12}{l^4} \le R_{abcd}R^{abcd} \le \frac{32}{l^4},
\ee
\be
\label{fieldstrengthtorus}
0\le F_{ab}F^{ab}= \frac{2 n^2}{V^2 (T^2+\tau_{\text{min}}^2)^2} \le \frac{2 n^2}{V^2 \tau_{\text{min}}^4}.
\ee

\subsection{Hyperbolic quotients}
One could also consider wormholes with compact 2d spatial cross sections $\mathcal{M}_2$ which are quotients of the hyperbolic plane by the fundamental groups (i.e., Kleinian groups) of those cross sections, which can have any genus $g>1$. One finds the solution with topology $\mathbb{R}\times\mathcal{M}_2$, which extrapolates nicely from the sphere and torus cases, as
\be
\label{newmetricthyperbola}
ds^2= \frac{1}{2 \pi \chi/V+T^2/l^2+2 \tau_{\min}^2/l^2} d T^2+(T^2+\tau_{\min}^2)(d \phi^2+\sinh^2\phi d \theta ^2 ),
\ee
\be
F= \frac{n}{V} \sinh{\phi}\, d\phi \wedge d\theta,
\ee
with $V$ the volume of $\mathcal{M}_2$, $\chi< 0$ depends on the topology of $\mathcal{M}_2$, and with
\be
\label{tauminsphere}
\frac{\tau_{\text{min}}^2}{l^2}= \sqrt{ \frac{\kappa n^2}{2 V^2 \alpha l^2 } +\left(\frac{\pi \chi}{V}\right)^2 }-\frac{\pi \chi}{V} \ge 1.
\ee
The behavior of the Kretschmann scalar is more complicated than in the other cases due to the competing curvature contributions, but remains nonsingular. The $U(1)$ field strength obeys \eqref{fieldstrengthtorus} (which is the same as \eqref{fieldstrengthsphere} with $V=4 \pi$).

\subsection{Truncated vs. two-sided vs. one-sided wormholes}

The one-sided solutions with minimal values of $\tau$ (the leftmost case in Figure \ref{wormholeshapes}) can be thought of as related to an amplitude for baby universe emission by an AdS$_3$ spacetime. As explained in e.g., \cite{Giddings:1988cx} such a process could produce a loss of quantum coherence in the bulk spacetime (due to entanglement with the baby universe).

The two-sided solutions found by defining the new coordinate $T$ and extending to negative values resemble the class of two-sided Euclidean wormholes which were first investigated in a holographic context in \cite{Maldacena:2004rf} and give rise to a factorization problem in the dual CFTs. This is shown in the middle image of Figure \ref{wormholeshapes}.

One can additionally exploit the isometries of AdS$_3$ to construct an approximation one-sided solution from the two-sided solution. On scales much larger than the wormhole mouth radius $\tau_{\text{min}}$, one has the full symmetry of AdS$_3$. One can identify one asymptotically AdS$_3$ with the other via an isometry. As long as the isometry is chosen so that the wormhole mouths end up widely separated, the result is an approximate saddle point which can contribute to the Euclidean path integral in the instanton gas approximation, as explained in \cite{Hebecker:2018ofv}. This is shown in the rightmost image of Figure \ref{wormholeshapes}. When these contribution are included and integrated, one is left with an effective action which is bilocal and therefore violates the cluster decomposition principle. Of course, quantum gravity may not respect exact cluster decomposition, but the associated boundary CFT \emph{must} and these wormhole contributions violate this as well \cite{Arkani-Hamed:2007cpn}.

If these wormholes contribute to the euclidean path integral, they are suggestive of violations of the unitarity of bulk spacetime, factorization, and/or cluster decomposition, even in the limit of small gravitational coupling. As explained in the introduction, the simplest evasion of these maladies would be for the solutions to be unstable, and therefore not valid saddle-point contributions. So we turn to evalauting this n the next section.

\section{Stability}
\label{stability}

In order for the solutions found in Section \ref{sec:solution} to contribute as saddle points in the classical $G \to 0$ limit, where they are most manifestly problematic they must also be stable (minimize the action). In this section we show that they are. We will assume the topology is $R \times \mathcal{M}_2$ for $\mathcal{M}_2$ any of the aforementioned compact spatial topologies.

\subsection{Degrees of freedom and SVT decomposition}
As is well known, there are no bulk gauge-invariant gravitational degrees of freedom. So we need only concern ourselves with perturbations of the $U(1)$ field and their gravitational backreactions. We may parameterize perturbations to $F$ as
\be
\label{pertF}
F= \frac{n}{V}dV+\epsilon dA,
\ee
where we have introduced $\epsilon$ for convenience to organize the order of perturbations throughout calculations. We may parameterize perturbations to $F$ this way despite the fact that for our spacetime not all closed forms are exact because $\int F$ is a conserved charge, the axion charge. We may choose the Coulomb gauge:
\be
A_0=\phi, \quad A_i=(A_1, A_2), \quad \nabla^i A_i=0.
\ee
Since $\phi$ is fixed by the Gauss's law constraint, we have only the divergences vector $A_i$ as a propagating degree of freedom.

As in standard cosmological perturbation theory, the spatial vectors of the SVT (scalar-vector-tensor) decomposition will only couple to other spatial vectors. So we only need to consider the vector sector of gravitational perturbations. There are two, a divergence free perturbation to the shift vector, and the vector component of the spatial metric tensor. We may choose the ``vector gauge'' \cite{JMStewart_1990} and eliminate the latter, which is convenient in this case since it leaves $\nabla^i$ unperturbed and leaves only the shift vector which manifestly non-dynamical. For the classical solutions, $N_i=0$, so we may write the perturbation as
\be
g_{0i}= \epsilon N_i.
\ee
Therefore we have two divergence-free vector perturbations to consider, one for the electromagnetic field and one for its gravitational backreaction:
\be
N_i=(N_1, N_2), \quad \nabla^i N_i=0, \quad A_i=(A_1, A_2), \quad \nabla^i A_i=0.
\ee
It will turn out that to be useful to treat the spatially\footnote{Technically in Euclidean signature all directions are spatial, here we mean to reference our choice of ADM slicing} homogeneous ``zero-modes'' separately from the spatially varying modes.

\subsection{Spatially varying modes}
The full calculation of the perturbative action to second order is contained in Appendix \ref{appB}. The upshot is that the electromagnetic and gravitational perturbations couple only via a single term (see \eqref{emperturb}):
\be
\sim \dot{A} \wedge N.
\ee
This term is identically 0 for all modes except the zero mode, however, by the following argument. First, we may further decompose these vector perturbations into eigenmodes of the (spatial) Laplace operator. Only perturbations of the same eigenmode will couple. So we may take:
\be
A_i=A_i^\lambda \phi^\lambda,
\ee
\be
N_i=N_i^\lambda \phi^\lambda,
\ee
where $\phi^\lambda$ a scalar function of spatial position which obeys the eigenvalue equation $\Delta \phi^\lambda= \phi^\lambda\lambda$ with $\Delta$ the spatial laplace operator, and where $A_i^\lambda$, $N_i^\lambda$ are independent of spatial position. Now recall that both $A_i$ and $N_i$ are divergence free, which translates to:
\be
A_i^\lambda \nabla^i \phi^\lambda=0 \to A_i^\lambda \sim \star \nabla_i \phi^\lambda,
\ee
\be
\dot{A}_i^\lambda \nabla^i \phi^\lambda=0 \to \dot{A}_i^\lambda \sim \star \nabla_i \phi^\lambda,
\ee
\be
N_i^\lambda \nabla^i \phi^\lambda=0 \to N_i^\lambda \sim \star \nabla_i \phi^\lambda,
\ee
where we use $\sim$ to mean ``up to a scalar function of the spacetime position'', that is they 
are colinear. The second step is only true because we are in $d=2$ and in the case $\lambda\ne0$. It then follows that:
\be
\dot{A}\wedge N=(\sim \star \phi^\lambda) \wedge (\sim \star \phi^\lambda) =0,
\ee
which means the two fields decouple, except for the zero mode $\lambda=0$.

Although this may seem lucky, it only appears so due to our desire to be as general as possible. In the case of the torus, for example, all what this means is that in the tangent space of a surface, two vectors transverse to the same wave vector must be parallel, so if their only coupling is via a cross product they do not couple. We have just generalized this argument to the other spaces. Since the action is local, this generalization to other global topologies is not surprising.

This additionally makes physical sense since for $A$ to represent a physical, gauge-invariant entity, it must induce a nontrivial connection on the $U(1)$ bundle. But any component of $A$ parallel to the shift vector $N$ is a coordinate transformation of a trivial connection. But only perturbations of the same wavevector will couple, so no physical perturbations will couple to the background in two dimensions.

Because the background decouples, we may then neglect the gravitiational field entirely since there is no backreaction and it includes no physical degrees of freedom of its own. We are left with the much simpler, decoupled, electromagnetic perturbation. Recalling \eqref{pertF}, we have
\be
F \wedge \star F = \frac{n^2}{V^2} dV \wedge \star dV+\epsilon\frac{n}{V}(dV \wedge \star dA+dA \wedge \star dV) + \epsilon^2 dA \wedge \star dA.
\ee
We see that the $\mathcal{O}(\epsilon^2)$ term is manifestly positive definite.

In conclusion, we have found that the action is a stable minimum for all perturbations with nonzero eigenvalues of the Laplacian for all topologies (even without needing to reference the boundary conditions). We now turn to the spatially homogeneous zero modes.

\subsection{Zero modes}
Applying the Poincar\'e--Hopf theorem and the Gauss--Bonnet theorem\footnote{The index of a vector field $v$ at its isolated zero $p$ is defined as follows: for a disk $D$ containing no other zeroes of $v$ than $p$, Index$_p(v)$ is the degree (winding number) of the map $u:\partial D\rightarrow S^1$ given by $u(z)=v(z)/|v(x)|$. The summation is over all isolated zeroes of $v$.} to $\mathcal{M}_2$, we get:
\be
\sum_{p\in\mathcal{M}_2: v(p)=0} \text{Index}_p(v)= \frac{1}{4\pi} \int_{\mathcal{M}_2} \sqrt{h}R_h.
\ee
This implies that any compact space of constant positive or negative curvature cannot permit a constant (zero mode) vector field (a generalization of the hairy ball theorem). So of the wormholes considered, there are no zero modes for the sphere or for hyperbolic quotients, and we need only concern ourselves with the case of the torus.

For the torus, the background is \eqref{TorusSolutionMetric} which has $N_0^2=(T^2/l^2+2 \tau_min^2/l^2)^{-1}$. We introduce the scale factor $a(T)^2=T^2+\tau_min^2$ for sake of notation. Our ansatz is
\be
N_i=(N_1(T), N_2(T)), \quad A_i=(A_1(t), A_2(t)),
\ee
with $\nabla^iN_i=\nabla^iA_i=0$ trivially since they have no spatial dependence. Using the results from Appendix \ref{appB} we find the action to be
\begin{align}
I= \int dT dV
&\left(
\left(\frac{n^2}{2 N_0 V^2 \alpha a^4}-\frac{4 \dot{a}^2}{N_0 \kappa a^2}\right)(N_1^2+N_2^2)\right.\\
\quad\quad&+\left.\left(\frac{1}{2 N_0 \alpha}\right)(\dot{A}_1^2+\dot{A}_2^2)+
\left(\frac{n}{N_0 V \alpha a^2}\right)( N_2 \dot{A_1}-N_1 \dot{A_2})
\right).
\end{align}
The shift vector is a Lagrange multiplier, as expected. Solving the corresponding equations of motion give the momentum constraints
\be
N_1= \frac{n V \kappa a^2}{n^2 \kappa - 8 V^2 \alpha a^2 \dot{a}^2} \dot{A}_2,
\ee
\be
N_2=-\frac{n V \kappa a^2}{n^2 \kappa - 8 V^2 \alpha a^2 \dot{a}^2} \dot{A}_1,
\ee
so the two perturbations are orthogonal as expected. The corresponding solutions for the $U(1)$ field are
\be
\dot{A}_i= c_i N_0 \alpha \left(1-\frac{n^2 \kappa}{8 V^2 a^2 \dot{a}^2}\right),
\ee
where $c_i$ are integration constants. The action for this solution comes out to be (simplifying it using the formula for $\tau_{\text{min}})$:
\be
I= \alpha \int dTdV \left(\left(
\frac{c_1^2+c_2^2}{2 N_0}\right)\left(1-\frac{\tau_{\text{min}}^4}{4 l^2 T^2}\right)
\right).
\ee
It is by no means obvious that this is postive definite, since there is a large negative contribution from deep within the wormhole throat $T\to 0$. However, we have not yet implemented the boundary conditions. As explained in \cite{Marolf_2006}, and which is important for interpreting the solutions, the boundary conditions for $D=3$ are to fix the field strength at the boundary, not (as may be expected) the surface potential. The condition may be written covariently as
\be
\label{boundaryconditions}
\lim_{T\to \pm \infty} \sqrt{h} n^{\mu} F_{\mu i}=constant,
\ee
where $n^\mu$ is a unit vector orthogonal to the boundary. For the fluctuations to preserve the boundary condition, the constant must be $=0$. Implementing this yields
\be
c_i/l=0.
\ee
And so these perturbations are forbidden by the boundary conditions. This means the torus too is (nontrivially) stable.

\subsection{Boundary Conditions and UV completions}
\label{BCandUV}

It is important to note that the boundary conditions \ref{boundaryconditions}, while the correct and only possible choice for $D=3$, are different from both the ``usual'' for a $U(1)$ gauge field and from those implemented in \cite{Loges:2022nuw, Hertog:2024nys}.

The boundary conditions in this work, fixing the \emph{normal} component of the field strength, correspond to fixing the $U(1)$ electric charge (by Gauss's law). As remarked on in $\cite{Marolf_2006}$, this is different than the more common choice of fixing the boundary potential, and this is significant because only in the latter case is the bulk gauge field dual to a CFT conserved charge. This matters for the proper CFT interpretation of these wormholes, e.g. which/which kind of operators may be expected to potentially exhibit related effects.

In \cite{Loges:2022nuw, Hertog:2024nys} the \emph{tangential} component of the gauge field is fixed. This corresponds to fixing the value of the axion charge, which in $D=3$ could be referred to as the dual photon charge. As explained in \cite{Marolf_2006} and more recently in \cite{Held:2026huj}, this is not allowed in $D=3$ and a simple way to see this is that the CFT operator dual to a scalar of mass $m$ in dimension $d$ is given by $\Delta_\pm = \frac{d}{2}\pm \sqrt{\frac{d^2}{4}+m^2 L^2}$ where +/- maps to Dirichlet/Neumann. For Neumann for $m=0$ (the usual axion case) the CFT would have an operator of dimension $\Delta_-=0$ which violates the unitarity bound. So this is not allowed.

Nevertheless, interestingly, for the case of the sphere which is the direct analogue of $\cite{Loges:2022nuw, Hertog:2024nys}$ (and the hyperbolic quotients) the stability did not depend in any way on the boundary conditions, meaning that even if one were to consider an (ultimately unphysical) toy model which is the direct analogue of their solutions, it would likewise exhibit the same stability. 

We can also consider the selection of boundary conditions from a bottom up perspective on the space of possible quantum gravity theories. From this point of view, one can consider all possible boundary conditions as candidate theories to be investigated, in which case the results of \cite{Hertog:2018kbz, Loges:2022nuw} could actually be viewed as complementary examinations of different theories. In our case, the results of \cite{Ishibashi:2004wx} show that the choice \ref{boundaryconditions} investigated by \cite{Marolf_2006} are the \emph{only} possible choice in $D=3$. So our results are themselves exhaustive in this sense. It is interesting to note that stability in our case does not depend on the boundary conditions except for the torus, and for the torus only for the zero mode. It would be interesting to know if this this pattern persists in $AdS_D$ for $D>3$ since it appears different than the pattern in $D=4$ Minkowski space found in the aforementioned references, where the boundary condition mattered decisively for e.g. the sphere.

Alternatively, one could attempt to take a top-down perspective on interpreting our stability results. Indeed, the setting of \cite{Arkani-Hamed:2007cpn} which connected axion wormholes to CFT cluster-decomposition, considered the embedding of axion wormholes in a toriodal compactification of Type IIB String Theory to $AdS_3$, and axions of some sort abound in String Theory compactifications  generally \cite{Svrcek:2006yi}. However, there are many barriers to UV embedding of our model. For example, one expects there to be dilaton fields coupled to any axions, and in some cases \cite{Astesiano} this has precluded embeddings while in others solutions remained and stability was retained \cite{Hertog:2024nys}. One also typically expects Chern-Simons terms for a gauge field (in the axion frame, this is related to the Peccei–Quinn theory cosine potential) or charged fields which could trigger the Polyakov confining mechanism \cite{Polyakov:1975rs}. It would be interesting to know if these obstacles can be overcome, or alternatively how each effect the existence and stability of solutions in generality, since this may offer answers to any associated wormhole paradoxes. Even more provocatively, the tensionless limit of string theory was investigated in $AdS_3$ and it was argued that including wormhole backgrounds was redundant and they should not be included as distinct saddles at all \cite{Eberhardt:2020bgq, Eberhardt:2021jvj}.

\section{Action}
\label{action}

The (classical) Euclidean action can be computed for each of the topologies considered and can be interpreted in light of Figure \ref{wormholeshapes}. For the first geometry, the action is related to the likelihood of baby universe emission, $\sim e^{- I}$. In the case of the second geometry, it provides a measure of free energy which can be compared with other solutions and can be used to determine the spectral form factor characterizing the appearance of a factorization problem. For the third geometry, one can compute sums over wormhole contributions to the path integral in an instanton gas approximation, which gives rise to nonlocal couplings with coupling constants dependent on the action. It is convenient to compute the action for the first geometry, then simply double it for the others (again, with the third geometry the action is correct in the sense of the instanton approximation).

To compute the action, one must add the boundary counterterm
\be
\label{boundaryaction1}
I_{ct}= \frac{1}{\kappa} \int \sqrt{h} (K-1/l),
\ee
where the first term is the Gibbons--Hawking--York term and the second term is from standard holographic renormalization. In the case where the boundary has $\chi\neq0$ and so is curved, one will additionally have a log divergence well known from CFT of the form
\be
\label{logdiv}
- \frac{c}{6} \log{(\Lambda/l)} \int \sqrt{h} R_h =
- 8 \pi^2 \chi (l/\kappa) \log{(\Lambda/l)},
\ee
which does indeed appear for our solutions as expected. We omit this term from the below, as well as any ``constant'' terms that do not depend only on $l$ since the are not renormalization independent given  \eqref{logdiv}.

\subsection{Torus}

The Euclidean action for the (half) wormhole solution \eqref{TorusSolutionMetric} using the ADM gravitational action \eqref{ADMaction2} and $U(1)$ action \eqref{emaction1} with boundary term \eqref{boundaryaction1} is
\be
\label{I0Torus}
I_{torus}^0= \frac{\pi V}{2} \frac{\tau_{\text{min}}^2}{\kappa l}= \frac{n \pi }{\sqrt{8 \alpha \kappa}},
\ee
where it turns out the gravitational and $U(1)$ actions contribute equally. Evidently the action is simply proportional to the number of units of magnetic charge, and it decreases with increasing gravitational of $U(1)$ coupling strength as one would expect.

\subsection{Sphere}
The Euclidean action for the (half) wormhole solution \eqref{sphereansatz} using the ADM gravitational action \eqref{ADMaction2} and $U(1)$ action \eqref{emaction1} with boundary term \eqref{boundaryaction1} with nonzero divergent term \eqref{logdiv} removed is
\be
I^0_{\text{sphere}}= \frac{8 \pi l}{\kappa} \left( 
\frac{\tau_{\text{min}}}{l} \sqrt{1+\frac{\tau_{\text{min}}^2}{l^2}} \arccot{ \frac{\tau_{\text{min}}}{\sqrt{l^2+\tau_{\text{min}}^2}}}+
\pi \log\left(1+2\frac{\tau_{\text{min}}^2}{l^2}\right)
\right).
\ee
The expression in terms of $n$ can be found using \eqref{tauminsphere}. Although it is much more complicated than \eqref{I0Torus}, for large $n$ it approaches a similar linear scaling relationship:
\be
\frac{dI^0_{\text{sphere}}}{dn}=\frac{\pi}{\sqrt{2\kappa \alpha}}.
\ee

\subsection{Hyperbolic quotients}
The action for the spatially hyperbolic wormholes depends on the specific quotient by a Kleinian group in question in an intricate way, so we do not pursue this calculation here. It is left for future development.

\section{Discussion}
\label{discussion}
In this paper we have found flat, positively curved, and negatively curved analogues of the Giddings--Strominger ``axion'' Euclidean wormholes in the context of asymptotically AdS$_3$ Einstein gravity with a $U(1)$ field, found that they are stable, and in some cases computed their classical action. We have done so by exploiting some convenient properties of the ADM formalism in the context of $D=3$ gravity. We have also found a convenient set of coordinates that cover the entirety of the two-sided solutions and which make manifest their nonsingularity. 

This builds on the recent work of \cite{Hertog:2018kbz,Loges:2022nuw, Hertog:2024nys} showing yet another class of stable wormhole contributions to the gravitational path integral that heighten the associated paradoxes (even as the allowed boundary conditions in $D=3$ mean our solutions differ in some important respects). It therefore seems increasingly likely that any hope that the problematic wormholes are simply unstable is unrealistic, however it is important to assess this in the context of a UV complete theory such as String Theory. Including dilaton fields, or better yet considering a construction such as those considered in \cite{Arkani-Hamed:2007cpn} and reexamining stability would provide stronger evidence. Some nice recent work along these lines has made progress \cite{Andriolo:2022rxc}

The full saddle point approximation should strictly speaking include $1$ loop contributions, such as were found in \cite{Giombi:2008vd}. These should in principle be obtained as integrals from local curvature invariants. It would be interesting to compute these in some cases. If it were, one could consider the full path integral, as in \cite{Maloney:2007ud} and examine corresponding pathologies and remedies systematically.

Even the classical action could be valuable for comparing to other solutions. In our two-sided case, we have not found a corresponding disconnected solution or any other solutions with the same boundary conditions. If one were to find any, one could look for a phase transition as a function of temperature or charge. Importantly, it must be remembered that in $D=3$ the boundary condition is fixed potential, not charge, as explained in \cite{Marolf_2006}.

Examining Euclidean AdS wormholes in other dimensions may also be interesting, to see if the stability argument can be generalized or to explore other applications such as some recent interesting proposals in AdS$_4$ \cite{Betzios:2024oli, Canfora:2025roy} and construction from 10d \cite{Loges:2023ypl}.

While this manuscript was undergoing revision, the very interesting, aforementioned, $\cite{Held:2026huj}$ also investigated $AdS_3$ axion wormholes. It would be very interesting to do a full comparison of results.

\appendix
\section{Axions, dual photons, and the Euclidean action}
Let $s=1$ for Euclidean signature and $s=-1$ for Lorentzian signature. Then the Hodge star operator $*$ when acting on a $p$-form in $D$ dimensions (spacetime) obeys: [see e.g., (2.83) in \cite{carroll2019spacetime}]
\be
*^2= (-1)^{p(D-p)} s.
\ee
This implies the following relation involving a $0$-form $\theta$ and a $p=(D-1)$-form $F$:
\be
\label{starlemma}
(F-*d\theta)\wedge * (F-*d\theta) = F \wedge * F + s d\theta \wedge *d\theta.
\ee
Now consider the action of such a $p$-form in a Euclidean or Lorentzian spacetime:
\be
s \int \frac{1}{2} F \wedge *F+\theta dF,
\ee
where $\theta$ is a 0-form Lagrange multiplier which enforces
\be
dF=0.
\ee
We can rewrite this using the relation \eqref{starlemma}:
\be
s \int \frac{1}{2} \big((F-*d\theta)\wedge * (F-*d\theta)-s d\theta \wedge *d\theta \big)+\theta dF.
\ee
One can do a field redefinition $\Tilde{F}=F-*d\theta$ and perfom the Gaussian integral over $\Tilde{F}$ to get
\be
\int \frac{1}{2} \big(- d\theta \wedge *d\theta \big)+s \int \theta dF.
\ee
In Lorentzian signature this is the correct sign, but in Euclidean signature the sign is reversed. Appropriate analysis of the boundary term gives that $\theta$ obeys Neumann boundary conditions.

The point is that the duality between between an axion and a $D-1$ form $dF$ does not commute with Wick rotation.

\section{Perturbative action}
\label{appB}
In this section we will assume that the topology of the spacetime is $\mathbb{R} \times \mathcal{M}_2$ where $\mathcal{M}_2$ is some compact and smooth surface and we will try to be as general as possible. We will use factors of $\epsilon$ to keep track of the order perturbations. We will use $a,b$ for spacetime indices and $i,j$ for purely spatial indices. Spacetime indices will be raised/lowered with $g_{ab}$ whereas spatial indices will be raised/lowered with $h_{ij}$ (which is a distinct operation in some cases).

The electromagnetic field ansatz is
\be
F= \frac{B}{V_{\Omega}} d \Omega +\epsilon dA,
\ee
\be
A=(0, A_1, A_2), \qquad \nabla^i A_i=0.
\ee
For the perturbations, we have chosen the Coulomb gauge and are looking only at the vector sector of the SVT decomposition since this is the only sector with a dynamical degree of freedom.

The gravitational field ansatz is
\be
g_{00}=N_0^2+\epsilon^2N^i N_j, \quad g_{0,i}=\epsilon N_i, \quad g_{ij}=h_{ij}=a(\tau)^2 \gamma_{ij}, \quad \nabla^i N_i=0,
\ee
where, echoing the above, the lapse function is $N^i=h^{ij}N_j\neq g^{ij}N_j$. In fact,
\be
g^{00}=\frac{1}{N_0^2}, \quad g^{0i}=-\epsilon \frac{N^i}{N_0^2}, \quad g^{ij}=h^{ij}+\epsilon^2 \frac{N^i N^j}{N_0^2},
\ee
where we are working in the ``vector gauge''. In this gauge the spatial metric is unperturbed, which is convenient because then $\nabla^i$ is not perturbed, and because $N_i$ is a Lagrange multiplier enforcing the momentum constraint. Again we are only looking at the vector sector of the SVT decomposition, and hence the shift vector is divergence-free just like the vector potential.

The electromagnetic action is
\be
I_{em}=\int d\tau d\Omega N_0 \sqrt{h} \frac{1}{4 \alpha} F_{ab} F^{ab}.
\ee
The nonzero components of $F$ are
\be
F_{0i}=-F_{i0}=\epsilon \dot{A}_i, \quad F_{12}=-F_{21}= \frac{B}{V} \sqrt{\gamma} + \epsilon ( \partial_1 A_2-\partial_2 A_1).
\ee

The gravitational action is
\be
I_g=-\frac{4 \pi \chi}{2 \kappa}\int d\tau N_0+\frac{1}{2 \kappa} \int d\tau d\Omega N_0 \sqrt{h}(K_{ab}K^{ab}-K^2+2\Lambda),
\ee
where
\be
K_{ij}=\frac{1}{2 N_0} \left( \dot{h}_{ij}-\epsilon (\nabla_i N_j+\nabla_j N_i)\right), \quad \Lambda = -\frac{1}{l^2},
\ee
and $\chi$ is the Euler characteristic of $\mathcal{M}_2$ and $l$ is the AdS radius. Note that the action involves raising $K_{ab}$ as a spacetime tensor even though it only has spatial components.

Our goal is to compute the action to $\mathcal{O}(\epsilon^2).$

First we write
\be
K_a^a=K_{ij}g^{ij}=\frac{2 H}{N_0}+\epsilon^2 \frac{H}{N_0} \frac{N^i N_i}{N_0^2},
\ee
where $H=\frac{\dot{a}}{a}$, so
\be
K^2=\frac{4 H^2}{N_0^2}\left(1+\epsilon^2 \frac{ N^i N_i}{N_0^2}\right).
\ee
Meanwhile,
\be
K_{ab}K^{ab}= \frac{2 H^2}{N_0^2} +\frac{1}{2 N_0^2} \nabla^i N^j \nabla_i N_j+\frac{2H^2}{N_0^2} N^i N_i,
\ee
so we have
\be
K_{ab}K^{ab}-K^2= -\frac{2 H^2}{N_0^2}(1+\epsilon ^2 N^i N_i)-\frac{\epsilon^2}{2 N_0^2} N^i \nabla^2 N_i,
\ee
where we integrated by parts to get the last term.

For a solution to the equation of motion, the first order perturbation should vanish. In this case it vanishes automatically. This make sense because the shift vector is the only perturbation and it is pure gauge.

Meanwhile for the electromagnetic field,
\be
\label{emperturb}
F_{ab}F^{ab}= \frac{B^2}{a^2 V^2}+\epsilon(2\partial^i A^j \epsilon_{ij})+\epsilon^2 \Big( 2(\partial^i A^j \epsilon_{ij})^2+
\frac{2}{N_0^2} \dot{A}_i \dot{A}^i+\sqrt{\gamma} \frac{4 B}{V} \dot{A}^i N^j \epsilon_{ij}+2 \frac{B^2}{V^2} a^2 \gamma N^i N_j \Big),
\ee
where the Levi--Civita symbol has $\epsilon_{12}=1$ and $\epsilon_{21}=-1$. Note that the $\epsilon$ term is a total derivative (Stokes' theorem) as expected.

\acknowledgments

We thank Andreas Karch and Thomas Van Riet for helpful conversation. 

H.-Y.S. was supported in part by the U.S. Department of Energy under Grant No. DE-SC0022021 and a grant from the Simons Foundation (Grant 651440, AK). This work was performed in part at
the Kavli Institute for Theoretical Physics (KITP), which is supported by NSF grant PHY-2309135.


\bibliographystyle{JHEP} 
\bibliography{refs}

\providecommand{\href}[2]{#2}\begingroup\raggedright\begin{thebibliography}{10}

\bibitem{Smolin:2005mq}
L.~Smolin, \emph{{The Case for background independence}},  \href{https://arxiv.org/abs/hep-th/0507235}{{\ttfamily hep-th/0507235}}.

\bibitem{HawkingPage}
S.W.~Hawking and D.N.~Page, \emph{{Thermodynamics of Black Holes in anti-De Sitter Space}}, \href{https://doi.org/10.1007/BF01208266}{\emph{Commun. Math. Phys.} {\bfseries 87} (1983) 577}.

\bibitem{Witten:1998zw}
E.~Witten, \emph{{Anti-de Sitter space, thermal phase transition, and confinement in gauge theories}}, \href{https://doi.org/10.4310/ATMP.1998.v2.n3.a3}{\emph{Adv. Theor. Math. Phys.} {\bfseries 2} (1998) 505} [\href{https://arxiv.org/abs/hep-th/9803131}{{\ttfamily hep-th/9803131}}].

\bibitem{Penington:2019kki}
G.~Penington, S.H.~Shenker, D.~Stanford and Z.~Yang, \emph{{Replica wormholes and the black hole interior}}, \href{https://doi.org/10.1007/JHEP03(2022)205}{\emph{JHEP} {\bfseries 03} (2022) 205} [\href{https://arxiv.org/abs/1911.11977}{{\ttfamily 1911.11977}}].

\bibitem{Almheiri:2019hni}
A.~Almheiri, R.~Mahajan, J.~Maldacena and Y.~Zhao, \emph{{The Page curve of Hawking radiation from semiclassical geometry}}, \href{https://doi.org/10.1007/JHEP03(2020)149}{\emph{JHEP} {\bfseries 03} (2020) 149} [\href{https://arxiv.org/abs/1908.10996}{{\ttfamily 1908.10996}}].

\bibitem{Rey:1998yx}
S.-J.~Rey, \emph{{Holographic principle and topology change in string theory}}, \href{https://doi.org/10.1088/0264-9381/16/7/102}{\emph{Class. Quant. Grav.} {\bfseries 16} (1999) L37} [\href{https://arxiv.org/abs/hep-th/9807241}{{\ttfamily hep-th/9807241}}].

\bibitem{Aspinwall:1993nu}
P.S.~Aspinwall, B.R.~Greene and D.R.~Morrison, \emph{{Calabi-Yau moduli space, mirror manifolds and space-time topology change in string theory}}, \href{https://doi.org/10.1016/0550-3213(94)90321-2}{\emph{Nucl. Phys. B} {\bfseries 416} (1994) 414} [\href{https://arxiv.org/abs/hep-th/9309097}{{\ttfamily hep-th/9309097}}].

\bibitem{Witten:1996qb}
E.~Witten, \emph{{Phase transitions in M theory and F theory}}, \href{https://doi.org/10.1016/0550-3213(96)00212-X}{\emph{Nucl. Phys. B} {\bfseries 471} (1996) 195} [\href{https://arxiv.org/abs/hep-th/9603150}{{\ttfamily hep-th/9603150}}].

\bibitem{Brandle:2002fa}
M.~Brandle and A.~Lukas, \emph{{Flop transitions in M theory cosmology}}, \href{https://doi.org/10.1103/PhysRevD.68.024030}{\emph{Phys. Rev. D} {\bfseries 68} (2003) 024030} [\href{https://arxiv.org/abs/hep-th/0212263}{{\ttfamily hep-th/0212263}}].

\bibitem{Stephens:1993an}
C.R.~Stephens, G.~'t~Hooft and B.F.~Whiting, \emph{{Black hole evaporation without information loss}}, \href{https://doi.org/10.1088/0264-9381/11/3/014}{\emph{Class. Quant. Grav.} {\bfseries 11} (1994) 621} [\href{https://arxiv.org/abs/gr-qc/9310006}{{\ttfamily gr-qc/9310006}}].

\bibitem{tHooft:1993dmi}
G.~'t~Hooft, \emph{{Dimensional reduction in quantum gravity}}, {\emph{Conf. Proc. C} {\bfseries 930308} (1993) 284} [\href{https://arxiv.org/abs/gr-qc/9310026}{{\ttfamily gr-qc/9310026}}].

\bibitem{Coleman:1988cy}
S.R.~Coleman, \emph{{Black holes as red herrings: Topological fluctuations and the loss of quantum coherence}}, \href{https://doi.org/10.1016/0550-3213(88)90110-1}{\emph{Nucl. Phys. B} {\bfseries 307} (1988) 867}.

\bibitem{Giddings:1988cx}
S.B.~Giddings and A.~Strominger, \emph{{Loss of incoherence and determination of coupling constants in quantum gravity}}, \href{https://doi.org/10.1016/0550-3213(88)90109-5}{\emph{Nucl. Phys. B} {\bfseries 307} (1988) 854}.

\bibitem{Giddings:1988wv}
S.B.~Giddings and A.~Strominger, \emph{{Baby Universes, Third Quantization and the Cosmological Constant}}, \href{https://doi.org/10.1016/0550-3213(89)90353-2}{\emph{Nucl. Phys. B} {\bfseries 321} (1989) 481}.

\bibitem{Hebecker:2018ofv}
A.~Hebecker, T.~Mikhail and P.~Soler, \emph{{Euclidean wormholes, baby universes, and their impact on particle physics and cosmology}}, \href{https://doi.org/10.3389/fspas.2018.00035}{\emph{Front. Astron. Space Sci.} {\bfseries 5} (2018) 35} [\href{https://arxiv.org/abs/1807.00824}{{\ttfamily 1807.00824}}].

\bibitem{Arkani-Hamed:2007cpn}
N.~Arkani-Hamed, J.~Orgera and J.~Polchinski, \emph{{Euclidean wormholes in string theory}}, \href{https://doi.org/10.1088/1126-6708/2007/12/018}{\emph{JHEP} {\bfseries 12} (2007) 018} [\href{https://arxiv.org/abs/0705.2768}{{\ttfamily 0705.2768}}].

\bibitem{Maldacena:2004rf}
J.M.~Maldacena and L.~Maoz, \emph{{Wormholes in AdS}}, \href{https://doi.org/10.1088/1126-6708/2004/02/053}{\emph{JHEP} {\bfseries 02} (2004) 053} [\href{https://arxiv.org/abs/hep-th/0401024}{{\ttfamily hep-th/0401024}}].

\bibitem{Witten:1999xp}
E.~Witten and S.-T.~Yau, \emph{{Connectedness of the boundary in the AdS / CFT correspondence}}, \href{https://doi.org/10.4310/ATMP.1999.v3.n6.a1}{\emph{Adv. Theor. Math. Phys.} {\bfseries 3} (1999) 1635} [\href{https://arxiv.org/abs/hep-th/9910245}{{\ttfamily hep-th/9910245}}].

\bibitem{Marolf:2020xie}
D.~Marolf and H.~Maxfield, \emph{{Transcending the ensemble: baby universes, spacetime wormholes, and the order and disorder of black hole information}}, \href{https://doi.org/10.1007/JHEP08(2020)044}{\emph{JHEP} {\bfseries 08} (2020) 044} [\href{https://arxiv.org/abs/2002.08950}{{\ttfamily 2002.08950}}].

\bibitem{Eberhardt:2021jvj}
L.~Eberhardt, \emph{{Summing over Geometries in String Theory}}, \href{https://doi.org/10.1007/JHEP05(2021)233}{\emph{JHEP} {\bfseries 05} (2021) 233} [\href{https://arxiv.org/abs/2102.12355}{{\ttfamily 2102.12355}}].

\bibitem{Saad:2021rcu}
P.~Saad, S.H.~Shenker, D.~Stanford and S.~Yao, \emph{{Wormholes without averaging}}, \href{https://doi.org/10.1007/JHEP09(2024)133}{\emph{JHEP} {\bfseries 09} (2024) 133} [\href{https://arxiv.org/abs/2103.16754}{{\ttfamily 2103.16754}}].

\bibitem{Saad:2021uzi}
P.~Saad, S.~Shenker and S.~Yao, \emph{{Comments on wormholes and factorization}},  \href{https://arxiv.org/abs/2107.13130}{{\ttfamily 2107.13130}}.

\bibitem{Iliesiu:2021are}
L.V.~Iliesiu, M.~Kologlu and G.J.~Turiaci, \emph{{Supersymmetric indices factorize}}, \href{https://doi.org/10.1007/JHEP05(2023)032}{\emph{JHEP} {\bfseries 05} (2023) 032} [\href{https://arxiv.org/abs/2107.09062}{{\ttfamily 2107.09062}}].

\bibitem{Schlenker:2022dyo}
J.-M.~Schlenker and E.~Witten, \emph{{No ensemble averaging below the black hole threshold}}, \href{https://doi.org/10.1007/JHEP07(2022)143}{\emph{JHEP} {\bfseries 07} (2022) 143} [\href{https://arxiv.org/abs/2202.01372}{{\ttfamily 2202.01372}}].

\bibitem{Hernandez-Cuenca:2024pey}
S.~Hern\'andez-Cuenca, \emph{{Wormholes and Factorization in Exact Effective Theory}},  \href{https://arxiv.org/abs/2404.10035}{{\ttfamily 2404.10035}}.

\bibitem{Hertog:2018kbz}
T.~Hertog, B.~Truijen and T.~Van~Riet, \emph{{Euclidean axion wormholes have multiple negative modes}}, \href{https://doi.org/10.1103/PhysRevLett.123.081302}{\emph{Phys. Rev. Lett.} {\bfseries 123} (2019) 081302} [\href{https://arxiv.org/abs/1811.12690}{{\ttfamily 1811.12690}}].

\bibitem{Giddings:1987cg}
S.B.~Giddings and A.~Strominger, \emph{{Axion Induced Topology Change in Quantum Gravity and String Theory}}, \href{https://doi.org/10.1016/0550-3213(88)90446-4}{\emph{Nucl. Phys. B} {\bfseries 306} (1988) 890}.

\bibitem{Loges:2022nuw}
G.J.~Loges, G.~Shiu and N.~Sudhir, \emph{{Complex saddles and Euclidean wormholes in the Lorentzian path integral}}, \href{https://doi.org/10.1007/JHEP08(2022)064}{\emph{JHEP} {\bfseries 08} (2022) 064} [\href{https://arxiv.org/abs/2203.01956}{{\ttfamily 2203.01956}}].

\bibitem{witten2026dualityaxionwormholes}
E.~Witten, \emph{Duality and axion wormholes},  2026.

\bibitem{Hertog:2024nys}
T.~Hertog, S.~Maenaut, B.~Missoni, R.~Tielemans and T.~Van~Riet, \emph{{Stability of axion-saxion wormholes}}, \href{https://doi.org/10.1007/JHEP11(2024)151}{\emph{JHEP} {\bfseries 11} (2024) 151} [\href{https://arxiv.org/abs/2405.02072}{{\ttfamily 2405.02072}}].

\bibitem{Maloney:2007ud}
A.~Maloney and E.~Witten, \emph{{Quantum Gravity Partition Functions in Three Dimensions}}, \href{https://doi.org/10.1007/JHEP02(2010)029}{\emph{JHEP} {\bfseries 02} (2010) 029} [\href{https://arxiv.org/abs/0712.0155}{{\ttfamily 0712.0155}}].

\bibitem{Castro:2011zq}
A.~Castro, M.R.~Gaberdiel, T.~Hartman, A.~Maloney and R.~Volpato, \emph{{The Gravity Dual of the Ising Model}}, \href{https://doi.org/10.1103/PhysRevD.85.024032}{\emph{Phys. Rev. D} {\bfseries 85} (2012) 024032} [\href{https://arxiv.org/abs/1111.1987}{{\ttfamily 1111.1987}}].

\bibitem{Jian:2019ubz}
C.-M.~Jian, A.W.W.~Ludwig, Z.-X.~Luo, H.-Y.~Sun and Z.~Wang, \emph{{Establishing strongly-coupled 3D AdS quantum gravity with Ising dual using all-genus partition functions}}, \href{https://doi.org/10.1007/JHEP10(2020)129}{\emph{JHEP} {\bfseries 10} (2020) 129} [\href{https://arxiv.org/abs/1907.06656}{{\ttfamily 1907.06656}}].

\bibitem{Karch:2020flx}
A.~Karch, Z.-X.~Luo and H.-Y.~Sun, \emph{{Holographic duality for Ising CFT with boundary}}, \href{https://doi.org/10.1007/JHEP04(2021)018}{\emph{JHEP} {\bfseries 04} (2021) 018} [\href{https://arxiv.org/abs/2012.02067}{{\ttfamily 2012.02067}}].

\bibitem{Cotler:2020ugk}
J.~Cotler and K.~Jensen, \emph{{AdS$_{3}$ gravity and random CFT}}, \href{https://doi.org/10.1007/JHEP04(2021)033}{\emph{JHEP} {\bfseries 04} (2021) 033} [\href{https://arxiv.org/abs/2006.08648}{{\ttfamily 2006.08648}}].

\bibitem{Cotler:2020hgz}
J.~Cotler and K.~Jensen, \emph{{AdS$_3$ wormholes from a modular bootstrap}}, \href{https://doi.org/10.1007/JHEP11(2020)058}{\emph{JHEP} {\bfseries 11} (2020) 058} [\href{https://arxiv.org/abs/2007.15653}{{\ttfamily 2007.15653}}].

\bibitem{Collier:2023fwi}
S.~Collier, L.~Eberhardt and M.~Zhang, \emph{{Solving 3d gravity with Virasoro TQFT}}, \href{https://doi.org/10.21468/SciPostPhys.15.4.151}{\emph{SciPost Phys.} {\bfseries 15} (2023) 151} [\href{https://arxiv.org/abs/2304.13650}{{\ttfamily 2304.13650}}].

\bibitem{Chandra:2022bqq}
J.~Chandra, S.~Collier, T.~Hartman and A.~Maloney, \emph{{Semiclassical 3D gravity as an average of large-c CFTs}}, \href{https://doi.org/10.1007/JHEP12(2022)069}{\emph{JHEP} {\bfseries 12} (2022) 069} [\href{https://arxiv.org/abs/2203.06511}{{\ttfamily 2203.06511}}].

\bibitem{Collier:2024mgv}
S.~Collier, L.~Eberhardt and M.~Zhang, \emph{{3d gravity from Virasoro TQFT: Holography, wormholes and knots}}, \href{https://doi.org/10.21468/SciPostPhys.17.5.134}{\emph{SciPost Phys.} {\bfseries 17} (2024) 134} [\href{https://arxiv.org/abs/2401.13900}{{\ttfamily 2401.13900}}].

\bibitem{Banados:1992gq}
M.~Banados, M.~Henneaux, C.~Teitelboim and J.~Zanelli, \emph{{Geometry of the (2+1) black hole}}, \href{https://doi.org/10.1103/PhysRevD.48.1506}{\emph{Phys. Rev. D} {\bfseries 48} (1993) 1506} [\href{https://arxiv.org/abs/gr-qc/9302012}{{\ttfamily gr-qc/9302012}}].

\bibitem{Mertens:2022irh}
T.G.~Mertens and G.J.~Turiaci, \emph{{Solvable models of quantum black holes: a review on Jackiw\textendash{}Teitelboim gravity}}, \href{https://doi.org/10.1007/s41114-023-00046-1}{\emph{Living Rev. Rel.} {\bfseries 26} (2023) 4} [\href{https://arxiv.org/abs/2210.10846}{{\ttfamily 2210.10846}}].

\bibitem{TongGaugeTheory}
D.~Tong, \emph{Lectures on gauge theory},  2018.

\bibitem{Marolf_2006}
D.~Marolf and S.F.~Ross, \emph{Boundary conditions and dualities: vector fields in {AdS}/{CFT}}, \href{https://doi.org/10.1088/1126-6708/2006/11/085}{\emph{Journal of High Energy Physics} {\bfseries 2006} (2006) 085}.

\bibitem{Barcelo:1995gz}
C.~Barcelo, L.J.~Garay, P.F.~Gonzalez-Diaz and G.A.~Mena~Marugan, \emph{{Asymptotically anti-de Sitter wormholes}}, \href{https://doi.org/10.1103/PhysRevD.53.3162}{\emph{Phys. Rev. D} {\bfseries 53} (1996) 3162} [\href{https://arxiv.org/abs/gr-qc/9510047}{{\ttfamily gr-qc/9510047}}].

\bibitem{Arnowitt:1959ah}
R.L.~Arnowitt, S.~Deser and C.W.~Misner, \emph{{Dynamical Structure and Definition of Energy in General Relativity}}, \href{https://doi.org/10.1103/PhysRev.116.1322}{\emph{Phys. Rev.} {\bfseries 116} (1959) 1322}.

\bibitem{JMStewart_1990}
J.M.~Stewart, \emph{Perturbations of friedmann-robertson-walker cosmological models}, \href{https://doi.org/10.1088/0264-9381/7/7/013}{\emph{Classical and Quantum Gravity} {\bfseries 7} (1990) 1169}.

\bibitem{Held:2026huj}
J.~Held, M.~Kaplan, D.~Marolf and Z.~Wang, \emph{{Axion Wormholes and the AdS/CFT Factorization Problem}},  \href{https://arxiv.org/abs/2601.02507}{{\ttfamily 2601.02507}}.

\bibitem{Ishibashi:2004wx}
A.~Ishibashi and R.M.~Wald, \emph{{Dynamics in nonglobally hyperbolic static space-times. 3. Anti-de Sitter space-time}}, \href{https://doi.org/10.1088/0264-9381/21/12/012}{\emph{Class. Quant. Grav.} {\bfseries 21} (2004) 2981} [\href{https://arxiv.org/abs/hep-th/0402184}{{\ttfamily hep-th/0402184}}].

\bibitem{Svrcek:2006yi}
P.~Svrcek and E.~Witten, \emph{{Axions In String Theory}}, \href{https://doi.org/10.1088/1126-6708/2006/06/051}{\emph{JHEP} {\bfseries 06} (2006) 051} [\href{https://arxiv.org/abs/hep-th/0605206}{{\ttfamily hep-th/0605206}}].

\bibitem{Astesiano}
D.~Astesiano, D.~Ruggeri, M.~Trigiante and T.~Van~Riet, \emph{Instantons and no wormholes in ${\mathrm{ads}}_{3}\ifmmode\times\else\texttimes\fi{}{S}^{3}\ifmmode\times\else\texttimes\fi{}{\mathrm{cy}}_{2}$}, \href{https://doi.org/10.1103/PhysRevD.105.086022}{\emph{Phys. Rev. D} {\bfseries 105} (2022) 086022}.

\bibitem{Polyakov:1975rs}
A.M.~Polyakov, \emph{{Compact Gauge Fields and the Infrared Catastrophe}}, \href{https://doi.org/10.1016/0370-2693(75)90162-8}{\emph{Phys. Lett. B} {\bfseries 59} (1975) 82}.

\bibitem{Eberhardt:2020bgq}
L.~Eberhardt, \emph{{Partition functions of the tensionless string}}, \href{https://doi.org/10.1007/JHEP03(2021)176}{\emph{JHEP} {\bfseries 03} (2021) 176} [\href{https://arxiv.org/abs/2008.07533}{{\ttfamily 2008.07533}}].

\bibitem{Andriolo:2022rxc}
S.~Andriolo, G.~Shiu, P.~Soler and T.~Van~Riet, \emph{{Axion wormholes with massive dilaton}}, \href{https://doi.org/10.1088/1361-6382/ac8fdc}{\emph{Class. Quant. Grav.} {\bfseries 39} (2022) 215014} [\href{https://arxiv.org/abs/2205.01119}{{\ttfamily 2205.01119}}].

\bibitem{Giombi:2008vd}
S.~Giombi, A.~Maloney and X.~Yin, \emph{{One-loop Partition Functions of 3D Gravity}}, \href{https://doi.org/10.1088/1126-6708/2008/08/007}{\emph{JHEP} {\bfseries 08} (2008) 007} [\href{https://arxiv.org/abs/0804.1773}{{\ttfamily 0804.1773}}].

\bibitem{Betzios:2024oli}
P.~Betzios and O.~Papadoulaki, \emph{{Inflationary Cosmology from Anti-de Sitter Wormholes}}, \href{https://doi.org/10.1103/PhysRevLett.133.021501}{\emph{Phys. Rev. Lett.} {\bfseries 133} (2024) 021501} [\href{https://arxiv.org/abs/2403.17046}{{\ttfamily 2403.17046}}].

\bibitem{Canfora:2025roy}
F.~Canfora, C.~Corral and B.~Diez, \emph{{Euclidean AdS wormholes and gravitational instantons in the Einstein-Skyrme theory}},  \href{https://arxiv.org/abs/2501.13024}{{\ttfamily 2501.13024}}.

\bibitem{Loges:2023ypl}
G.J.~Loges, G.~Shiu and T.~Van~Riet, \emph{{A 10d construction of Euclidean axion wormholes in flat and AdS space}}, \href{https://doi.org/10.1007/JHEP06(2023)079}{\emph{JHEP} {\bfseries 06} (2023) 079} [\href{https://arxiv.org/abs/2302.03688}{{\ttfamily 2302.03688}}].

\bibitem{carroll2019spacetime}
S.M.~Carroll, \emph{{Spacetime and Geometry}: {An Introduction to General Relativity}}, Cambridge University Press (7, 2019), \href{https://doi.org/10.1017/9781108770385}{10.1017/9781108770385}.

\end{thebibliography}\endgroup
\end{document}